\documentclass[a4paper,11pt]{article}
\usepackage{jcappub}

\makeatletter
\newcommand\footnoteref[1]{\protected@xdef\@thefnmark{\ref{#1}}\@footnotemark}
\makeatother

\usepackage{amsfonts,amsmath,amssymb,bm,tensor}
\usepackage{comment}
\usepackage{ifpdf}
\usepackage{slashed}
\usepackage{braket}
\usepackage[mathscr]{eucal}
\usepackage{cancel}  
\usepackage{multirow}

\makeatother

\usepackage{color}
\definecolor{darkred}{rgb}{0.6,0,0}
\definecolor{darkgreen}{rgb}{0.992447,0.623778,0.034597}
\definecolor{ppink}{rgb}{1,0.4,0.4}
\definecolor{bblue}{rgb}{0.284602,0.317763,0.963947}
\definecolor{mygreen}{rgb}{0,0.7,0}
\definecolor{myred}{rgb}{1,0.3,0.4}
\definecolor{myblue}{rgb}{0.2,0.3,1}

	\newcommand{\eV}{\mathrm{eV}}

\newcommand{\bs}{\boldsymbol}
\newcommand{\tx}{\text}

\newcommand{\qcq}{\quad,\quad}

\newcommand{\df}{\text{d}}
\newcommand{\p}{\partial}

\begin{document}


\title{
     Oscillons of Axion-Like Particle:
     \\
     Mass distribution and power spectrum
}

\author[a,b]{Masahiro Kawasaki,}
\author[c]{Wakutaka Nakano,}
\author[a]{Hiromasa Nakatsuka,}
\author[a]{and Eisuke Sonomoto}

\affiliation[a]{ICRR, University of Tokyo, Kashiwa, 277-8582, Japan}
\affiliation[b]{Kavli IPMU (WPI), UTIAS, University of Tokyo, Kashiwa, 277-8583, Japan}
\affiliation[c]{Department of Physics, University of Tokyo, Tokyo, 113-0033, Japan}

\emailAdd{kawasaki@icrr.u-tokyo.ac.jp}
\emailAdd{m156077@icrr.u-tokyo.ac.jp}
\emailAdd{sonomoto@icrr.u-tokyo.ac.jp}
\emailAdd{hiromasa@icrr.u-tokyo.ac.jp}

\abstract{
    In string theory,
    the simultaneous existence of many Axion-Like Particles (ALPs) are suggested
    over a vast mass range,
    and a variety of potentials have been developed in the context of inflation.
    In such potentials shallower than quadratic, 
    the prominent instability can produce localized dense objects, 
    oscillons.
    Because of the approximate conservation of their adiabatic invariant,
    oscillons generally survive quite long,
    maybe up to the current age of the universe in the case of ultra-light ALPs with $m \sim 10^{-22}\ {\rm eV}$.
    Such oscillons can have significant effects on the evolution of the recent universe.
    
    In this paper, 
    we investigate the oscillons of the pure-natural type potential 
    by classical lattice simulation 
    to explore the key quantities necessary for phenomenological application:
    the number density of oscillons, the oscillon mass distribution, the energy ratio of oscillons to the ALP field,
    and the power spectrum.
    Then, 
    we evolve these values
    in consideration of the analytic decay rate.
}

\maketitle

\section{Introduction}
\label{sec_intro}

Although the $\Lambda$-CDM cosmology excellently explains the various observations,
some exceptions are found in small scale physics, so-called "small scale crisis"~\cite{Weinberg:2013aya} 
such as the core-cusp problem~\cite{Moore:1999gc,deBlok:2001hbg,Oh:2008ww,amorisco2012dark}.
One of the solutions to the small scale crisis is the fuzzy dark matter, 
the ultra-light bosonic dark matter with mass $\sim 10^{-22}\ \eV$.
Such extremely light mass is naturally explained by the pseudo-scalar field whose mass is prohibited by a shift symmetry.
Since the string theory predicts a large number of axion-like particles (ALPs) with a broad range of masses, the string axion is one of the promising candidates of the ultralight bosonic fields.

In the literature, the ALP potential is often simply assumed as the cosine type in analogy with QCD axion.
However, other types of potentials are suggested in the context of inflation~\cite{Dong:2010in,Kallosh:2013hoa,Kallosh:2013tua,Kallosh:2013yoa,Silverstein:2008sg,McAllister:2008hb,Nomura:2017ehb,Dubovsky:2011tu}.
These potential are shallower than quadratic, 
which results in the prominent instability when ALP oscillates around the potential minimum.
Such instability can produce the gravitational waves in the reach of future observations~\cite{Soda:2017dsu,Kitajima:2018zco}.

The large instability also produces the localized soliton of ALP, oscillon~\cite{Bogolyubsky:1976yu,Gleiser:1993pt,Copeland:1995fq}.
For the QCD axion, such soliton is sometimes called "axiton"~\cite{Vaquero:2018tib,Kolb:1993hw,Kolb:1993zz}.
The formation of oscillons is confirmed on pure-natural type potential numerically~\cite{Amin:2011hj,Kawasaki:2019czd,Hong:2017ooe}.
When the sizable energy fraction of the ALPs is confined inside the oscillons, 
the constraints on the homogeneous ALPs~\cite{Li:2013nal,Porayko:2018sfa,Nori:2018pka} should be reconsidered.
The lifetime of oscillons is intensively studied~\cite{Olle:2019kbo,Ibe:2019vyo,Zhang:2020bec} in the last few years, 
and it is found that oscillons live quite long.
The long lifetime opens new possibilities of oscillons discussed in several literature \cite{Olle:2019kbo,Amin:2020vja,Buckley:2020fmh,Prabhu:2020pzm}.
However, no research has been done so far on the current abundance of ALP oscillons 
including the effects of their long lifetime and decay.

In this paper, 
we study properties of oscillons produced from ALP by lattice simulation.
First, we calculate the mass distribution, the energy ratio, the number density, and the average mass of oscillons after the formation.
Using these values,
we compare the simulated density power spectrum to the analytic formula extended from the Poisson distribution.
Finally, we study their time evolution by estimating the analytic decay rate of oscillons.

Sec.~\ref{sec_oscillontheory} explains the nature of oscillons in the pure natural type potential.
Sec.~\ref{sec_setup} describes the setup of lattice simulations,
which derive the number density of oscillons, the energy ratio of oscillons to ALP, initial field dependence, mass distribution, and the power spectrum
in Sec.~\ref{sec_simulation}.
Sec.~\ref{sec_timedep} discusses the decay of oscillons
and our paper concludes in Sec.~\ref{sec_conclusion}.

\section{Oscillons of Axion-Like Particle}
\label{sec_oscillontheory}

Various kinds of ALP potentials are derived
when the instanton effect breaks the shift symmetry of ALP.
In this paper,
we use the pure natural type potential arising from the coupling 
between the ALP and the pure Yang-Mills gauge fields~\cite{Nomura:2017ehb},
\begin{align}
    V(\phi) 
    = \frac{m^2F^2}{2p}
	\left[ 
	  1 - \left(1 + \frac{\phi^2}{F^2}\right)^{-p}
	\right]
	.
\label{eq_potential}
\end{align}
This potential is flatter than quadratic for $p > -1$, 
which causes the instability of the field
and produces localized objects, oscillons~\cite{Hong:2017ooe,Kawasaki:2019czd}.

We focus on the ALP which starts to oscillate during the radiation dominated era.
In this potential,
the relation between the Hubble parameter at the onset of the oscillation
and the initial field value $\phi_i$ is represented by~\cite{Kawasaki:2011pd}
\begin{align}
	\left( \frac{H}{m} \right)^2
    \simeq  \frac{1}{m^2\phi}\frac{\p V(\phi)}{\p  \phi}  ~\bigg|_{\phi= \phi_i}
	= \left( 1+\frac{\phi_i^2}{F^2} \right)^{-(p+1)}.
	\label{eq:initial_Hubble}
\end{align}
Thus, the large $|\phi_i|$ and large $p$ delay the start of the oscillation,
and the ALP mass must be 
\begin{align}
    m 
    \gtrsim 
    10^{-28}\ {\rm eV}\ 
    \left( 1 + \frac{\phi_i^2}{F^2} \right)^{\frac{p+1}{2}}
    \left( \frac{H_{\rm eq}}{10^{-28}\ {\rm eV}} \right).
\end{align}
where $H_{\rm eq}$ is the Hubble parameter at the matter radiation equality.

\subsection{Instability Analysis}
\label{sec_instability_analysis}

In this subsection, 
we will show that the coherent oscillation in this potential 
causes two kinds of instabilities~\cite{Kitajima:2018zco},
which affect oscillon formation as we will see in Sec.~\ref{sec_initialdependence}.

We perform the linear analysis by decomposing the field into the background field $\phi_0(t)$ and its perturbation $\delta\phi(x)$ as
\begin{align}
	\phi(t,\bs x) = \phi_0(t) + \delta\phi(t,\bm x).
\end{align}
We assume that the background field coherently oscillates as 
$ \phi_0(t)=\phi_i (a_i/a)^{3/2} \cos mt$
where the lower index $i$ represents the beginning of the oscillation,
and define the Fourier components as $\phi_k$ 
where $k$ is the comoving wavenumber.

The first kind of instability is the parametric resonance 
which causes oscillon formation.
When $\phi \ll F$,
we can approximate the potential as $ V_{,\phi\phi}/m^2 \simeq 1 +  \lambda  (\phi/F)^2 /2$ 
with $\lambda = -6(1+p)$.
In this case, the time evolution of $\phi_k$ is described by the Mathieu equation,
\begin{align}
    \frac{d^2}{d (mt)^2} \phi_k  
	& +[A_k -2 q \cos(2mt)] \phi_k =0 ,
	\\
    q & = - \frac{\lambda}{8} \left(\frac{\phi_i}{F} \right)^2  \left( \frac{a_i}{a} \right)^3,\\
    A_k &= 1+\left( \frac{k}{am} \right)^2 - 2q.
\end{align}
Neglecting the cosmic expansion ($a=a_i$), 
the most efficient resonance band is
\begin{align}
	\frac{k}{a_i m} 
	\sim \sqrt{2q} 
	= \frac{\sqrt{|\lambda|} |\phi_i|}{2F}.
\end{align}
The oscillon formation becomes effective
when the perturbations with the wavelength of the oscillon size 
are amplified by the parametric resonance.
If these instabilities amplify the perturbations up to $\mathcal O(1)$,
the localized field configuration, the oscillon, can be formed,

The second kind of instability is tachyonic instability, 
which hinders the coherent oscillation needed for the oscillon formation.
The tachyonic instability is caused by the negative effective mass,
\begin{align}
    V,_{\phi\phi} < 0
	\quad \Leftrightarrow \quad
	\frac{\phi}{F} 
	> \frac{1}{\sqrt{1+2p}}
	\equiv \frac{\phi_c}{F}. 
\end{align}
Thus, the potential has the inflection point only when $p > -1/2$.
The tachyonic instability occurs in the small $k$ mode with  
\begin{align}
	\frac{k}{am} < \left| \frac{V_{,\phi\phi}}{m^2} \right|^{1/2} .
\end{align}

\subsection{Oscillon Profile}
\label{sec_oscillon_profile}

The oscillon configuration is quasi-stable 
because of the approximate conservation of the adiabatic invariant~\cite{Kasuya:2002zs,Kawasaki:2015vga},
\begin{align}
  I \equiv 
    \frac{1}{\omega}
	\int \df ^3 \bm x \overline{\dot{\phi}^2},
\end{align}
where $\omega$ is the frequency of the periodic motion
and the overline represents the time average over one period.
When we assume that $I$ is invariant and set $\phi(t,\bm x) = \Phi(r) \cos(\omega t)$,
the lowest energy configuration of the scalar field $\Phi(r)$ is determined by the following differential equation:
\begin{align}
	\frac{\df^2 \Phi}{\df r^2} 
	+\frac{2}{r} \frac{\df \Phi}{\df r}
	+\omega^2 \Phi
	-2 \, \overline{\frac{\df V}{\df \Phi}(\Phi \cos(\omega t)) \cos(\omega t)}
	=0
	,
	\label{eq_oscillonsEq}
\end{align}
with boundary conditions $\Phi|_{r\to \infty} =0$ and $\df \Phi/\df r|_{r\to 0} = 0$.
The profile of produced oscillons depends on the parameter $\omega$.
When we set $\omega$ smaller, the initial condition of $\Phi (r=0)$ becomes larger, i.e., we obtain a larger oscillon.
The oscillon is produced under the condition that the Eq.~\eqref{eq_oscillonsEq} has the solution, 
and the potential Eq.~\eqref{eq_potential} satisfies this condition only when $p > -1$ as we mentioned earlier.

The range of $\omega$ is constrained by the stability condition
\begin{align}
    \frac{dI}{d\omega} < 0 .
    \label{eq:omega_conditin}
\end{align}
This charge $I$ relates to the classical particle number conservation~\cite{Vak:1973st,Lee:1991ax,Mukaida:2016hwd,Mukaida:2014oza}.

\subsection{Oscillon Decay}
\label{sec_oscillon_decay}

Because the adiabatic invariance is only approximate,
oscillons gradually decay by emitting the self-radiation.
In this subsection, 
we calculate the classical decay rate of the oscillon by perturbative method following refs.~\cite{Ibe:2019vyo,Zhang:2020bec}.

The classical decay rate is defined by
\begin{align}
	\Gamma(M_\tx{osc})\equiv \frac{|\overline{\dot M_\tx{osc}}|}{M_\tx{osc}} ,
	\label{eq_decayrate_def}
\end{align}
where $M_\tx{osc}$ is the energy of the oscillon configuration.
We take the radiation term $\xi (t, r)$ as
\begin{align}
    \phi (t, {\bf x}) = \Phi(r) \cos(\omega t) + \xi (t, r).
\end{align}
The linear approximated equation of motion for $\xi (t, r)$ is 
\begin{align}
    \left[ \, \square + V''(\Phi \cos(\omega t)) \right] \xi = -V' (\Phi \cos(\omega t)) + 2 \, \overline{V' (\Phi \cos(\omega t)) \cos (\omega t)} \cos (\omega t) ,
\end{align}
where prime denotes the derivative of $\phi$.

It is very convenient to use the Fourier expansions in time for the calculation of the decay rate.
We expand $\xi (t, r) = \sum_{j=3} \xi_j (r) \cos (j \omega t)$ and the potential terms.
The equation of motion for $\xi_j$ is written as
\begin{align}
   \left[ \nabla^2 + \kappa_j^2 \right] \xi_j = - \mathcal{S}_j ,
\end{align}
where $\kappa_j = \sqrt{j^2 \omega^2 - m^2}$ and $\mathcal{S}_j (r)$ is the source term which includes the combination of $\xi_k$'s.
The solution of $\xi_j$'s is derived under the same boundary condition as for the oscillon profile calculation,
but it includes not only inhomogeneous solution but also homogeneous solution, i.e. free field solution.
We compare the derived solutions with the formal outgoing solutions at the origin to remove the homogeneous solution,
\begin{align}
    \xi_j (0) = \int_0^\infty dr' \mathcal{S}_j (r') \, r' \cos (\kappa_j r') .
\end{align}

The radiative solution of $\xi$ for $r \to \infty$ is obtained as
\begin{align}
    \xi_j^{\mathrm{rad}} (t, r) \approx & \frac{1}{4 \pi r} \sum_{j=3} \tilde{\mathcal{S}}_j (\kappa_j) \cos (\kappa_j r - j \omega t) , \\
    \tilde{\mathcal{S}}_j (k) = & \int d^3 {\bf r}' \mathcal{S}_j (r') \, e^{-i {\bf k \cdot r}'} .
\end{align}
The time averaged classical energy loss rate is given by
\begin{align}
    \overline{\dot M_\tx{osc}} = 4 \pi r^2 \, \overline{\partial_t \xi^{\mathrm{rad}} (t, r) \partial_r \xi^{\mathrm{rad}} (t, r)} = - \frac{1}{8 \pi} \sum_{j=3} j \omega \kappa_j \tilde{\mathcal{S}}_j (\kappa_j)^2 .
    \label{eq_decayrate}
\end{align}
Therefore, we can write $\Gamma = \sum_{j=3} \Gamma_j$ and find $\Gamma_j \ge 0$.
In the following, we just calculate $\Gamma_{3}$ and assume $\Gamma\simeq \Gamma_3$ because there is relation of $|\xi_j| \gg |\xi_{j+1}|$ (hence $\Gamma_j \gg \Gamma_{j+1}$) confirmed in $\omega \gtrsim 0.8 m$ except for
poles of $\Gamma_3$ according to the previous study~\cite{Zhang:2020bec}.

\begin{figure}
	\centering
	\includegraphics[width=.80\textwidth]{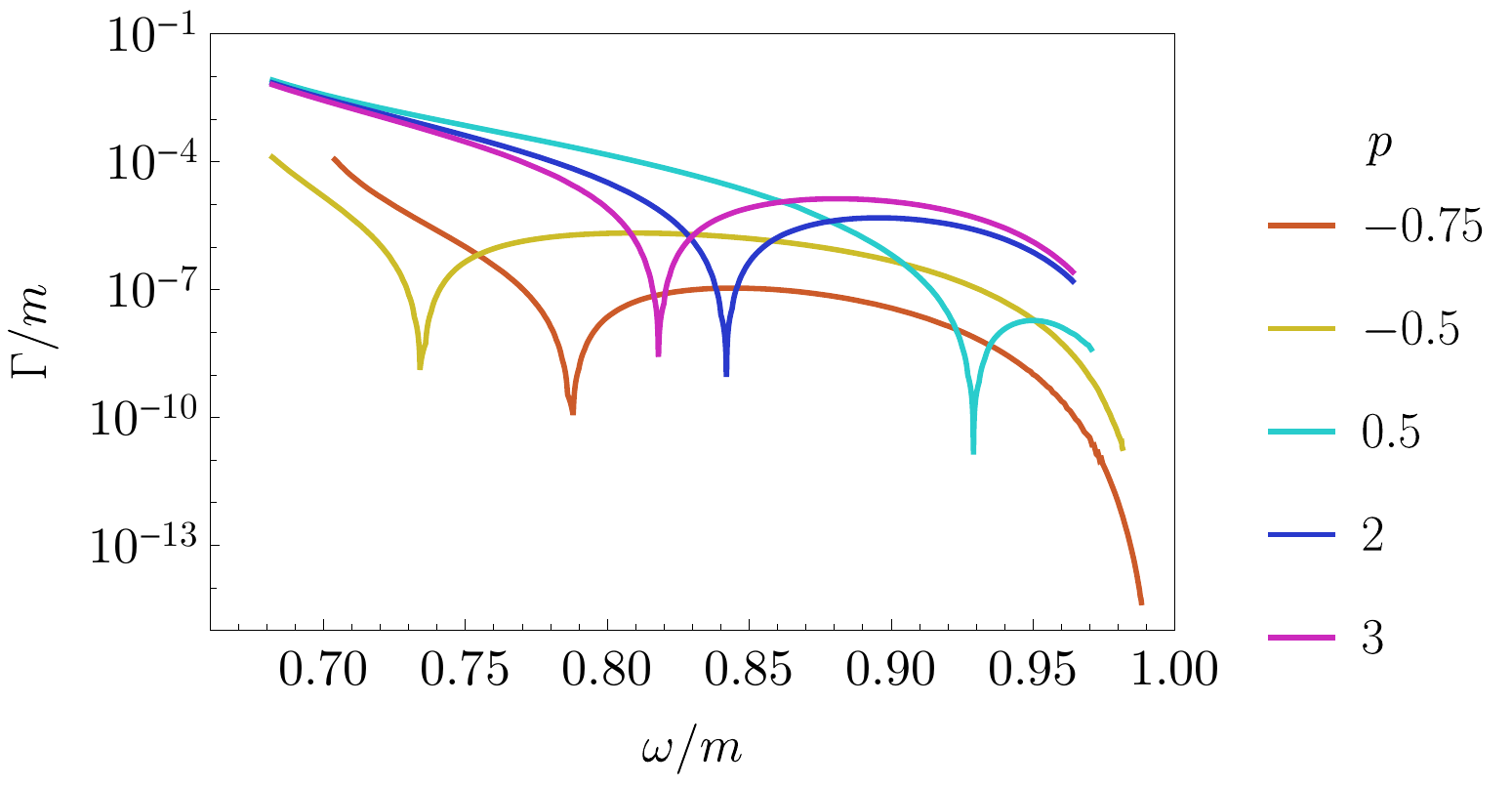}
	\caption{
    Decay rates of oscillons calculated based on Eqs.~\eqref{eq_decayrate_def} and \eqref{eq_decayrate} with $j=3$.
    The oscillon mass is related to  $\omega$.
    The different lines shows the different $p$.
    }
	\label{fig_decay}
\end{figure}

The decay rates are shown in Fig.~\ref{fig_decay} as a function of $\omega$. 
The upper limit of $\omega$ is determined by the relation $dI/d\omega = 0$.
In Fig.~\ref{fig_decay},
there are poles that the decay rate is extremely small.
Generally, the position of poles of the decay rate $\Gamma_j$ does not correspond to that of $\Gamma_{j+1}$ and therefore, 
the relation $\Gamma_{j+1} \gg \Gamma_{j}$ could hold at the poles of $\Gamma_j$.
This means that the lifetime at poles calculated here is the maximal lifetime, 
i.e., the lifetime could be overestimated by the pole effect.

The time dependence of oscillon mass is formally given by
\begin{align}
    M_\tx{osc}(t) = M_\tx{osc}(0) \exp\left( -\int^t_0 \Gamma(M_\tx{osc}(t'))\df t' \right).
    \label{eq_oscillontimedepFormula}
\end{align}
We will discuss the time evolution of oscillons 
based on Eq.~\eqref{eq_oscillontimedepFormula} in Sec.~\ref{sec_timedep}.

\subsection{Extended Poisson Distribution}
\label{sec_poisson}

The dark matter oscillons produce large isocurvature fluctuations.
The simple approximation assuming the random distribution of point particles with a single mass 
leads to Poisson distribution. 
In this section,
extending it with consideration of broad mass of oscillons 
and energy conservation inside the horizon scale,
we analytically derive the power spectrum of ALP oscillons.
In the following, we use the comoving coordinate $\bs x$ and momentum $\bs k$.

Assuming that 
the spatial profile of oscillons is approximated by the Gaussian function 
and their velocity is almost zero due to the Hubble friction,
the energy density of a single oscillon $\rho_{\rm osc}$ with the total energy $M$ is represented by
\begin{align}
	\rho_\tx{osc}(t,\bs x; \bs x_0, M) 
	&\simeq
	\frac{[\omega(M)\psi_0(M)]^2 }{2} 
	\exp\left(
	-\frac{a(t)^2 (\bs x-\bs x_0)^2}{R(M)^2}
	\right)
	\label{eq_assumedprofile}
\end{align}
where $\bs x_0$, $R(M)$, $\omega(M)$, and $\psi_0(M)$ are 
the comoving position, the physical radius, the oscillation frequency, and the central field value of the oscillon.
Then, the total energy of the oscillon is written as
\begin{align}
	M=a(t)^3 \int \df^3 \bs x\rho_\tx{osc}(t,\bs x; \bs x_0, M) 
	= 
	\frac{[\omega(M) \psi_0(M)]^2 }{2}
	(\pi R(M)^2)^{3/2} .
\end{align}

The density perturbation $\delta_\text{tot} $ is given by the contribution of $N$ oscillons over the average energy density of ALP ${\rho}_\phi$ including oscillons, i.e, 
\begin{align}
	\delta_{\rm tot}(\bm{x})
	&\equiv
	\sum_{i}^N 
	\frac{\rho_{\rm osc}(t,\bs x; \bs x_i, M_i)}{\rho_\phi(t)},
\end{align}
where  and
$\bs x_i$ and $M_i$ is the comoving coordinate and mass of the $i$-th oscillon.
Then, the Fourier component of $\delta_{\rm tot}$ is
\begin{align}
	\delta_{\bs k} (t)
	&\equiv
	\int \df \boldsymbol x^3 \delta_{\rm tot}(\boldsymbol x) e^{-i\bs x\cdot \bs k}, 
	\\
	&=
	\frac{1}{a(t)^3\rho_\phi(t)}  \sum_{i}^N  e^{-i\bs x_i\cdot\bs k}
    M_i \exp\left( -\frac{R(M_i)^2 k^2}{4a(t)^2}  \right),
	\label{eq_oscillonmodes}
\end{align}

To calculate the sum $\sum_i^N$,
let us consider the comoving volume $V$ including $N$ oscillons.
When the volume $V$ is smaller than the horizon size at oscillon formation,
we expect that the oscillons are randomly distributed.
The power spectrum is given by the ensemble average of the oscillon coordinates,
\footnote{
    Here, the power spectrum is defined as
    \begin{align}
        \braket{\delta_{\bs k}\delta_{\bs k'}}
        &= 
        (2\pi)^3 
        \delta^3(\bs k-\bs k')
        P_{\tx{osc},k} 
        \quad,\quad
        \delta_{\bs k} \equiv
        \int\df x^3 
        \delta_{\rm tot} e^{-i\bs x\cdot \bs k}.
    \end{align}
}
\begin{align}
	P_{\tx{osc},k}(t) 
	&\equiv  
	\left(
    	\prod_i \int_V \frac{\df \boldsymbol x_i^3}{V}
	\right)
	\frac{|\delta_{\bs k} (t)|^2}{V},
	\nonumber
\\	&=
	\left( \frac{1}{a(t)^3\rho_\phi(t)} \right)^2
	\left(
    	\prod_i \int_V \frac{\df \boldsymbol x_i^3}{V}
	\right)
	\frac{1}{V}
	\sum_{l,n}^N  e^{i(\bs x_l-\boldsymbol x_n)\cdot\bs k}
    M_lM_n \exp\left( -\frac{(R(M_l)^2+R(M_n)^2) k^2}{4a(t)^2}  \right)
	\nonumber
\\&	\simeq
	\left( \frac{1}{a(t)^3\rho_\phi(t)} \right)^2
	\frac{1}{V}
	\sum_{i}^N
	M_i^2 \exp\left( -\frac{R(M_i)^2 k^2}{2a(t)^2}  \right),
	\nonumber
\\&	=
	 \frac{r_\tx{osc}^2}{n_\tx{osc}(t) } 
    \frac{
        \Braket{
    M^2 \exp\left( -\frac{R(M)^2 k^2}{2a(t)^2}  \right)
        }
    }{\braket{M}^2}
    \label{eq_powerspec_pre}
	 .
\end{align}
where we define $\braket{}$ as the average over mass $M$, and $n_{\rm osc}$ and $r_{\rm osc}$ are number density and energy ratio of oscillons to the ALP, respectively:
\footnote{
    We have used physical energy density and comoving number density.
}
\begin{align}
    n_{\rm osc} \equiv  \frac{N}{V}
\qcq
    r_\tx{osc}  = \frac{n_\tx{osc}(t) \braket{M}}{a(t)^3 \rho_\phi(t)}
    \quad.
\end{align}

On a scale larger than the horizon size at the oscillon formation,
the position of oscillons is not purely randomized
because of the energy conservation inside the horizon. 
More precisely, energy is conserved in a volume inside which the energy transfer of the scalar field is not effective.
Thus, the power spectrum damps over the scale of energy transfer at the oscillon formation $k_c$, 
which is smaller than the horizon size.
Considering the energy conservation, 
the average of oscillon coordinates in Eq.~\eqref{eq_powerspec_pre} is modified, and we introduce the following damping factor (see Appendix~\ref{app_energyConserve} for the derivation)
\begin{align}
	P_{\tx{osc},k}(t) 
	&=
	\frac{r_\tx{osc}^2}{n_\tx{osc}(t) } 
    \frac{
        \Braket{
    M^2 \exp\left( -\frac{R(M)^2 k^2}{2a(t)^2}  \right)
        }
    }{\braket{M}^2}
	 K(k/k_c), 
	\label{eq_oscillonPowerSpec}
    \\ 
	K(x)
	&\equiv 
	\left[
	1 - \left( \frac{2}{x} \right)^2\sin^2\left( \frac{x}{2} \right)
	\right]. 
	\label{eq_oscillonPowerSpecFactor}
\end{align}
where $k_c$ is the cutoff scale of perturbations
and should be determined by simulation.

Note that the above formula reproduces the Poisson distribution when we neglect the size of oscillons ($R(M)\to 0$), use the monochromatic mass ($\braket{M^2}=\braket{M}^2 $) and ignore the energy conservation ($k_c/k\ll 1$).

We call the above formula as extended Poisson distribution in this paper.
In the following section, 
we will confirm that the above estimation is in good agreement with the power spectrum derived from the numerical simulation.

\section{Simulation Setup}
\label{sec_setup}

In this section, 
we briefly explain the setup of the lattice simulations
to investigate the oscillon formation on various $p$'s and initial field values.
We use tilde to denote a simulation value normalized by a physical scale.
The normalized length and time in units of $m$ 
and the field value in the units of $F$ are given by
\begin{align}
(\tilde t ,\tilde{ \bs x}) \equiv m(t,\bs x)
\qcq
\tilde{\phi} \equiv \frac{\phi}{F}.
\end{align}
In this normalization, the mass of oscillon $M$ is normalized as
\begin{align}
    \tilde M = \frac{m}{F^2} M.
\end{align}

\subsection{Simulation Parameters}

We take the following initial condition in simulations:
\begin{align}
	\tilde \phi _i (\bm x)&=\tilde \phi_i (1+\xi(\bm x) )
	\qcq
	\tilde\phi_i' = 0
	\qcq
	H_i = m^{-1}
	\qcq
	\tilde a_i =1,
\end{align}
where $\tilde \phi_i $ is order one, and $\xi(\bm x)$ is the initial noise defined by the scale-free power spectrum
\begin{align}
	\braket{\xi_{\bm k}\xi_{\bm k'}}
	=(2\pi)^3 \delta^3(\bs k-\bm k') \frac{2\pi^2}{k^3 } \mathcal P_\xi ,
\end{align}
with a small constant $\mathcal P_\xi=2.1\times 10^{-9}$ as a reference value\footnote{
    Note that this definition slightly changes in two dimension as
	\[\braket{\xi_{\bm k}\xi_{\bm k'}}
	= (2\pi)^2 \delta^2(\bs k-\bm k') \frac{2\pi}{k^2} \mathcal P_\xi.
    \]
}.
Because we assume the radiation-dominated era 
and use the conformal time $\tilde \tau \equiv \sqrt{2\tilde t} = \tilde{a}$,
the initial physical and conformal times are given by 
$\tilde{t}_{i} = m/(2H_i)=1/2$ and $\tilde \tau_i = m\tau_i = 1$.

The parameters of our simulations are summarized 
in Table~\ref{table_latticeparam_3d} for three dimensions 
and Table~\ref{table_latticeparam_2d} for two dimensions.
The small box setup ($L=4,8$) is used to investigate the initial field dependence in Sec.~\ref{sec_initialdependence}, 
and the large box setup ($L=16,32$) is used to investigate the oscillon mass distribution in Sec.~\ref{sec_massdist}.
Both setup satisfies the condition that 
all oscillons should be resolved at the end of simulations.
We use the threshold value $\rho_{\rm th}$ to identify the oscillons explained in the next subsection.
\begingroup
\renewcommand{\arraystretch}{1.3}
\begin{table}[t]
	\caption{
	    The setups of our three dimensional numerical simulations.
	    We use the setup with the small box size when we need a large number of iterations 
	    and the large one when we calculate the oscillon mass distribution 
	    and the power spectrum. 
	}
	\vspace{0.2cm}
	\centering
	\begin{tabular}{|c||c|c|c|c|c|} 
	    \hline
		\multirow{2}{*}{$p$} & Box size & Grid size &    Time step   &   Final time     &     Threshold \\
     	                     &   $L$    &    $N$    & d$\tilde \tau$ & $\tilde{\tau}_e$ & $\tilde{\rho}_{\rm th}$ \\ 
     	\hline \hline
		$-0.75$ & $8$  & $512^3$  & $4.0\times 10^{-3}$ & $81.0$ & \multirow{3}{*}{$0.05$} \\ 
		\cline{1-5}
		\multirow{2}{*}{$-0.5$} & $4$  & $256^3$  & $4.0\times 10^{-3}$ & $121.0$ & \\ 
	                            & $32$ & $1024^3$ & $8.0\times 10^{-3}$ & $81.0$  & \\ 
		\hline
		$0.5$ & $4$ & $512^3$ & $2.0\times 10^{-3}$ & $101.0$ & $0.04$ \\ 
		\hline
		\multirow{2}{*}{$2$} & $4$  & $256^3$  & \multirow{3}{*}{$4.0\times 10^{-3}$} & $121.0$ & \multirow{3}{*}{$0.01$} \\ 
	                         & $16$ & $1024^3$ &                                      & $81.0$  & \\ 
		\cline{1-3}\cline{5-5}
	    $3$  & $4$ & $256^3$ & & $121.0$ & \\
		\hline
	\end{tabular}
	\label{table_latticeparam_3d}
\end{table}
\endgroup
\begingroup
\renewcommand{\arraystretch}{1.3}
\begin{table}[t]
	\caption{
	    The setup of our two dimensional numerical simulations.
	    We use this setup only for the calculation of the oscillon power spectrum.
	}
	\vspace{0.2cm}
	\centering
	\begin{tabular}{|c||c|c|c|c|c|} 
	    \hline
		\multirow{2}{*}{$p$} & Box size & Grid size &    Time step   &   Final time     &     Threshold \\
     	                     &   $L$    &    $N$    & d$\tilde \tau$ & $\tilde{\tau}_e$ & $\tilde{\rho}_{\rm th}$ \\ 
     	\hline \hline
	    $-0.5$ & $1024$ & $32768^2$ & $8.0\times 10^{-3}$ & $81.0$ & 0.025 \\
		\hline
	\end{tabular}
	\label{table_latticeparam_2d}
\end{table}
\endgroup

The lattice code adopts the fourth-order symplectic integration scheme for time evolution 
and the fourth-order central differential scheme for spatial derivatives with the periodic boundary condition~\cite{Ibe:2019vyo,Ibe:2019lzv,Kawasaki:2019czd}.
We investigated the effect of the boundary condition with different box sizes 
and confirmed that the box size difference does not change the results.

\subsection{Oscillon Identification}

The numerical simulation of localized object are investigated in the early works for Q-ball \cite{Hiramatsu:2010dx} and oscillon \cite{Amin:2010dc,Amin:2019ums,Lozanov:2019ylm}.

To estimate the contribution from the tail of the oscillon profile correctly,
we regard the connected regions with the energy density
larger than the threshold value $\tilde \rho_\text{th}$ 
and one more grid outside the surface of them as oscillons.
The threshold value $\tilde{\rho}_{\rm th}$ is chosen so that
$\tilde \rho_{\rm th} =\alpha \tilde \rho_\phi(\tilde \tau_{\rm e})$
where $\alpha$ is $\mathcal{O}(0.1)$ parameter 
and the time $\tilde \tau_{\rm e}$ is the end of oscillon formation.

The mass of the oscillons is estimated 
by summing over the energy density $\tilde{\rho}$ inside the identified region~\cite{Hiramatsu:2010dx} as
\begin{align}
	\tilde{M}_{\rm osc} 
	= 
	\sum_{\bs x \in V_\tx{osc}} \tilde \rho(\bs x)
	\left( \frac{\tilde{a}L}{N} \right)^d,
\end{align}
where $d$ is the spatial dimension
and the sum is over the grid $\bs x$ inside the oscillon volume $V_\tx{osc}$.
The number of oscillon inside the box $N_{\rm osc}$ is given 
by the number of these disconnected regions.

We define the energy ratio of oscillons to the total scalar field as
\begin{align}
	r_{\rm osc}(t)
	= \frac{\sum_i^{N_{\rm osc}} \tilde{M}_{{\rm osc}, i}}{\tilde{\rho}(t) (\tilde{a}(t) L)^d},
	\label{eq_def_rosc}
\end{align} 
and the normalized comoving number density of the oscillons $\tilde{n}_{\rm osc}$ 
and the average mass of the oscillon $\langle \tilde{M}_{\rm osc} \rangle$ as
\begin{align}
	\tilde n_{\rm osc} = \frac{N_\tx{osc}}{L^d}, \quad
	\braket{\tilde M_\tx{osc}}  = \frac{\sum^{N_\tx{osc}}_i \tilde M_{{\rm osc},i}}{N_{\rm osc}}.
	\label{eq_normalizedNum}
\end{align}
Because of the contribution from the surface region,
the threshold value and the spatial resolution slightly change $M_{\rm osc}$ and hence $r_{\rm osc}$.
We found that the estimation of energy ratio defined in Eq.~\eqref{eq_def_rosc} differs 
less than $5\%$ even if we change the spatial resolution and the threshold value up to a factor of ~$2$.

\section{Simulation Results}
\label{sec_simulation}

In this section, we exhibit the results of the classical lattice simulations.
In Sec.~\ref{sec_energyratio},
we show the time dependence of the number of oscillons and $r_\tx{osc}$.
In Sec.~\ref{sec_initialdependence},
we discuss the initial value dependence of $r_{\rm osc}$
changing the initial field value $\tilde\phi_i$ with small box setup for each $p$.
In Sec.~\ref{sec_massdist}, 
we simulate the oscillon formation with the large box setup to investigate the oscillon mass distribution.
Finally in Sec.~\ref{sec_powerspec}, 
we numerically calculate the power spectrum of oscillons and compare it with the analytic formula in Eq.~\eqref{eq_oscillonPowerSpec}.

\subsection{Time Evolution of Oscillon Number and its Energy Ratio}
\label{sec_energyratio}

\begin{figure}[t]
	\begin{center}
    	\begin{tabular}{c}
    		\begin{minipage}{0.48\hsize}
    			\begin{center}
    				\includegraphics[width=1.0\textwidth]{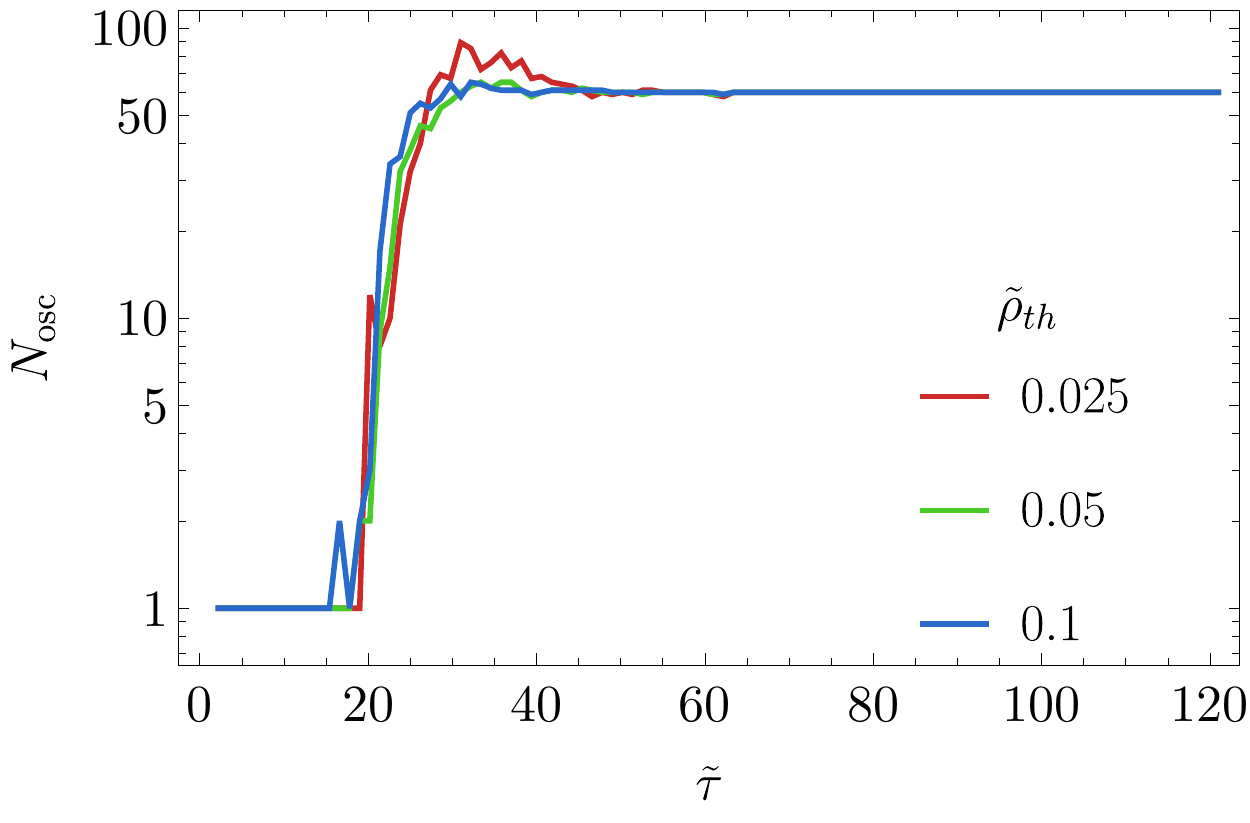}
    			\end{center}
    		\end{minipage}
    			
    		\begin{minipage}{0.48\hsize}
    			\begin{center}
    			    \includegraphics[width=1.0\textwidth ]{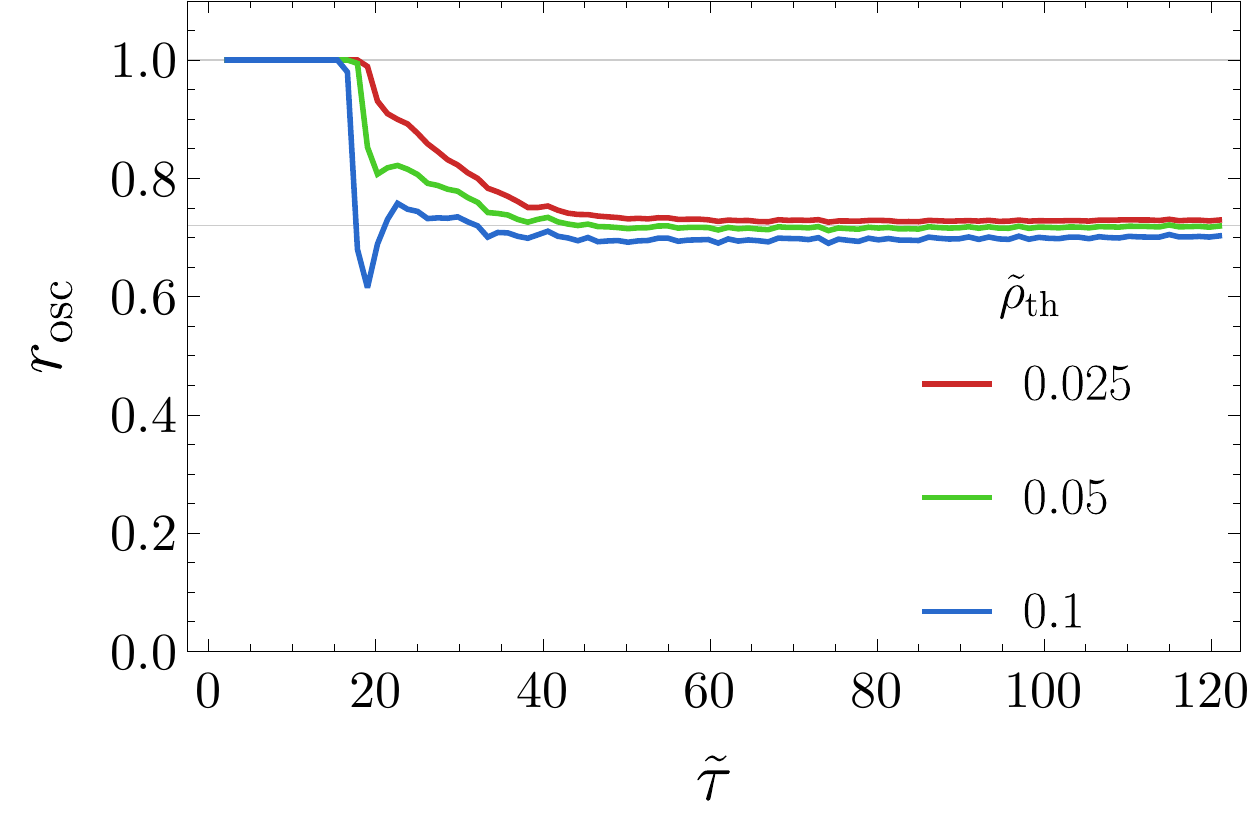}
    			\end{center}
    		\end{minipage}
    	\end{tabular}
		\caption{
		    The time evolution of the number (left figure) 
		    and energy ratio (right figure) of oscillons. 
		    The energy ratio converge to the stable value at $\tilde\tau >40$, 
		    which we call the energy ratio of oscillons $r_\tx{osc}$. 
		}
		\label{fig_typicaltimedependence}
	\end{center}
\end{figure}

First, 
we simulate the time evolution of the oscillon number $N_{\rm osc}$ and the oscillon energy ratio $r_{\rm osc}$ 
changing the threshold value $\tilde{\rho}_{\rm th}$ to observe their threshold dependence
during the oscillon formation.
The result for $p=-0.5$ and $\tilde{\phi}_i = 5.0\pi$ with $L=4$ 
is shown in Fig.~\ref{fig_typicaltimedependence}.
The red, green, and blue lines represent $\Bar{\rho}_{\tx{th}}=0.025,~0.05,~0.1$ respectively.

The left figure shows the number of oscillons.
After the formation around $\tilde\tau\sim 20$, 
$N_\tx{osc}$ slightly decreases until $\tilde \tau\sim 40$
because of the merger of the produced oscillons and the damping of fluctuations larger than $\tilde{\rho}_{\rm th}$ by the gradient term and the cosmic expansion.
Then, $N_\tx{osc}$ becomes stable independent of the threshold values by the simulation end. 
In the following calculation, 
we evaluate the energy of each oscillon when $N_\tx{osc}$ becomes stable.

The right figure shows the time evolution of $r_\tx{osc}(t)$.
After the oscillon number becomes stable, 
the energy ratio also becomes stable, 
which hardly depends on the threshold values.
Hereafter, we call this stable value as the energy ratio of oscillon $r_\tx{osc}$
\footnote{
    For some choice of parameters, 
    $r_\tx{osc}(t)$ slightly decreases 
    since the large oscillons decay in the simulation time.
    We discuss the details of decay in the next section.
}.

\subsection{The Initial Field Value Dependence of Energy Ratio}
\label{sec_initialdependence}

Next, we examine the dependence of $r_\tx{osc}$ on the initial field value $\phi_i$.
For each index $p$,
we perform the simulations with the small box setup in Table \ref{table_latticeparam_3d}
changing initial field value $\phi_i$.

\begin{figure}
	\centering
	\includegraphics[width=.70\textwidth ]{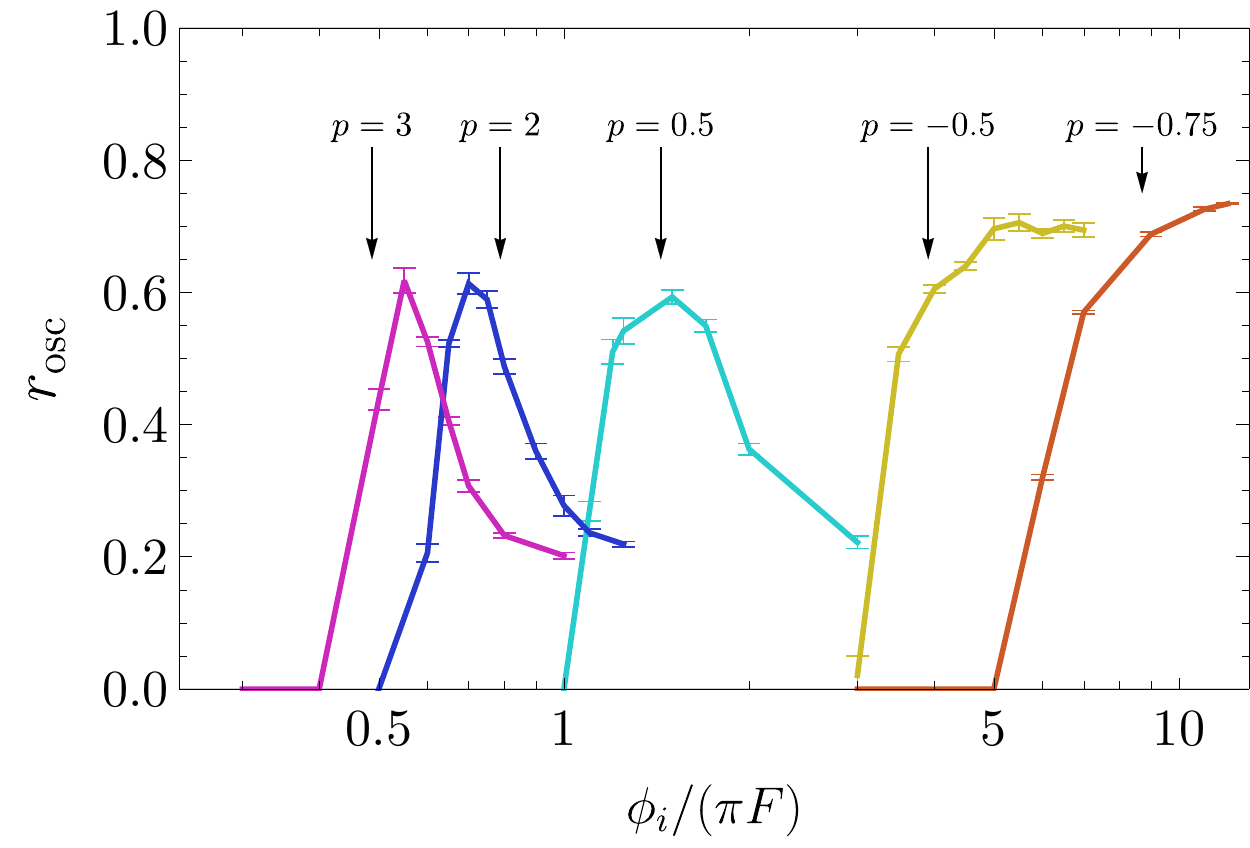}
	\caption{
		The oscillon energy ratio with different index $p$ and initial field value $\phi_i$.
		The different line corresponds to the different index $p$.
		The error bar shows the variance over 
		three times of simulations with different random seeds.
	}
	\label{fig_roWithDiffPhi0}
\end{figure}

We show $r_\tx{osc}$ as a function of $\phi_i$ for each $p$ in Fig.~\ref{fig_roWithDiffPhi0}.
For a small $\phi_i$, 
because the instability is too small for the oscillon formation,
$r_{\rm osc}$ ends up in zero.
For a larger $\phi_i$, $r_{\rm osc}$ also gets larger 
because the parametric resonance becomes effective.
Then, for $p = -0.75, -0.5$, $r_{\rm osc}$ converges on a plateau in Fig.~\ref{fig_roWithDiffPhi0}, and we define  $\tilde{\phi}_{i,m}$ as the typical field value on the plateau.
On the other hand, 
when $p > -0.5$,
the tachyonic instability amplifies the modes different from the oscillon size,
which hinders the subsequent oscillon formation as discussed in Sec.~\ref{sec_instability_analysis}.
Therefore, 
the oscillon formation rate has a peak at some value of $\phi_i$ for $p > -0.5$ 
and we define it as $\tilde \phi_{i,m}$
\footnote{
To confirm the effect of tachyonic instability, 
we calculate $r_\tx{osc}$ for $p=2$ with smaller initial noise $\mathcal P_\xi$.
We find that $r_\tx{osc}$ increases for $\phi_i=0.9\pi>\phi_{i,m}$ and decreases for $\phi_i=\phi_{i,m}$.
For the former case, the small initial noise delays the tachyonic instability.
For the latter case, $\phi_{i,m}$ is small enough to avoid the tachyonic instability and the small noise just results in small oscillon formation.
This result is consistent with the effect of tachyonic instability.
}.
For the later use,
we perform the same analysis in two dimensions with $p=-0.5$.

We comment on the initial noise dependence of the oscillon energy ratio. 
Because the instability growth only logarithmically depends on the size of the initial fluctuations,
we expect that $r_\tx{osc}$ does not depend on the initial noise much.
Actually,
we confirmed that $r_\tx{osc}$ typically varies less than $10\%$
when we change initial noise $\mathcal P_\xi$ by a factor of about $10$.

\subsection{The Mass Distribution of Oscillons}
\label{sec_massdist}

Here, we fix the initial field value as $\tilde \phi_i = \tilde \phi_{i,m}$ derived in the previous subsection
and perform simulations for $p=-0.5$ and $2$ in the large box setup 
and $p=-0.75,~0.5$, and $~3$ in the small box setup 
in Table \ref{table_latticeparam_3d}.
We have simulated in each setup three times with different random seeds to reduce the stochastic noise.

\begin{figure}
	\centering
	\begin{tabular}{c}
		\begin{minipage}{0.47\hsize}
			\begin{center}
			\includegraphics[width=1.0\textwidth ]{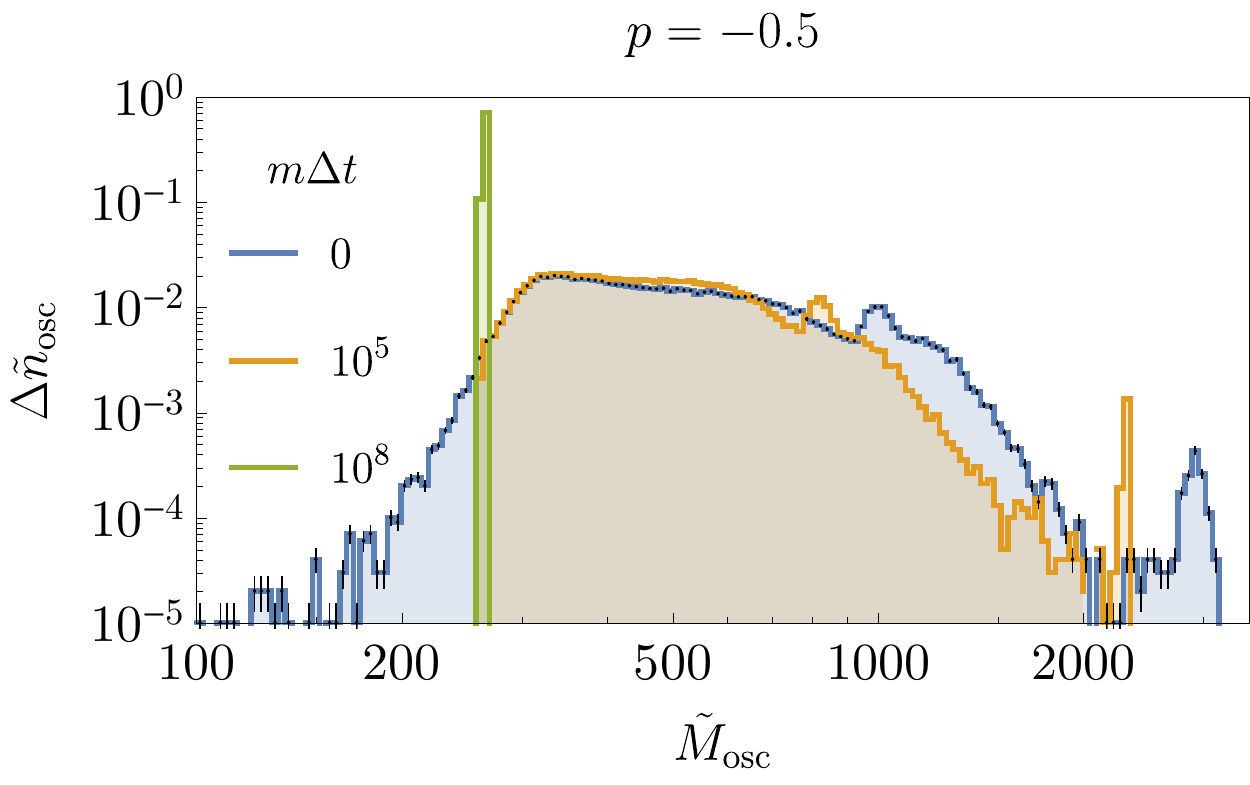}
			\end{center}
		\end{minipage}
		
		\begin{minipage}{0.45\hsize}
			\begin{center}
			\includegraphics[width=1.0\textwidth ]{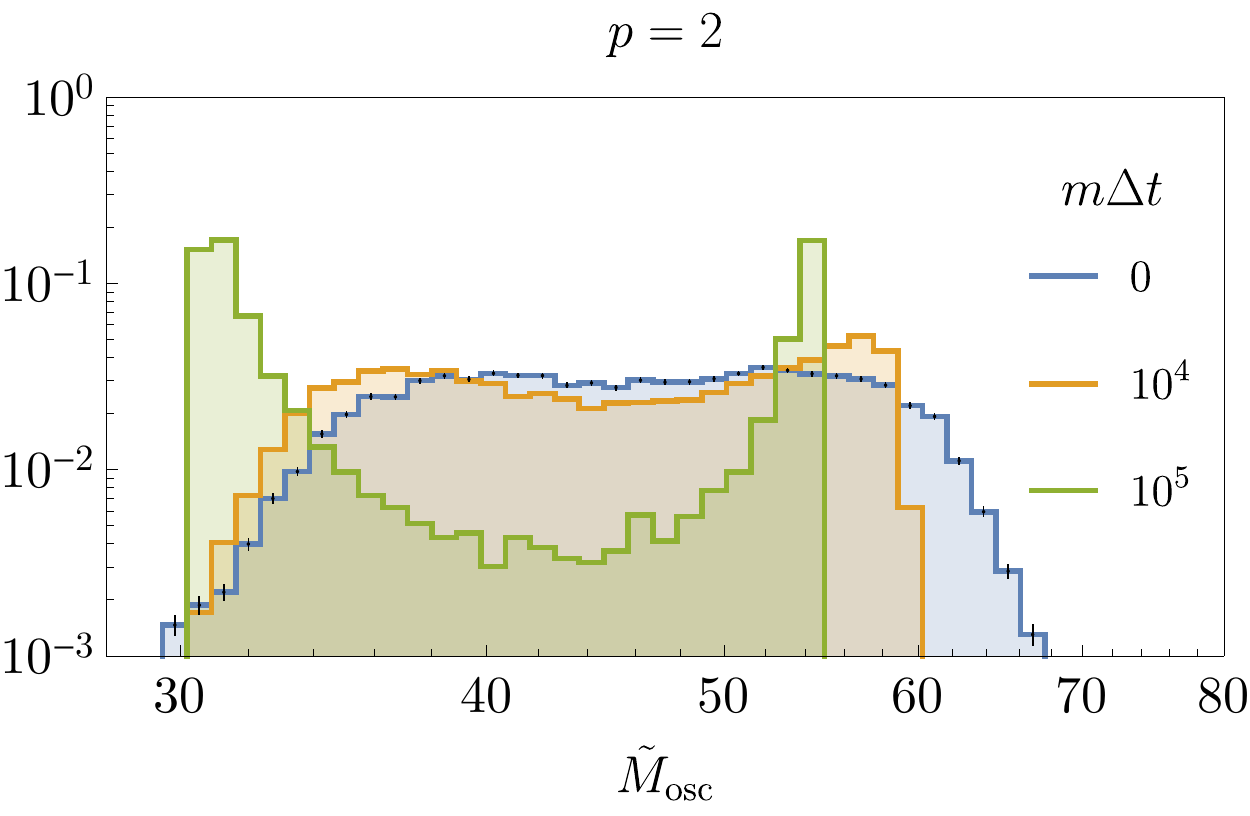}
			\end{center}
		\end{minipage}
	\end{tabular}
	\caption{
		The histogram of oscillon mass with $p=-0.5$ (left figure) and $p=2$ (right figure).
		The horizontal axis shows the normalized oscillon mass $\tilde{M}_{\rm osc}$
		and the vertical axis shows the comoving number density in each bin $\Delta \tilde{n}_i \equiv N_i/L^3$ 
		where $N_i$ is the number of oscillons in the $i$-th bin.
		Each bar is binned by the log scale $[10^{0.01 i},10^{0.01 (i+1)}]$ and
		the error bar is given by $\sqrt{N_i}/L^3$.
		The blue region shows the oscillon distribution at $\tilde \tau = 81$
		and the orange and green regions show the oscillon distribution 
		after the oscillon decay over $m\Delta t$ 
		discussed in Sec.~\ref{sec_timedep}.
		The split form of $p=2$ with $m\Delta t= 10^5$ comes from the pole structure of the decay rate, which has the uncertainty discussed in Sec.~\ref{sec_oscillon_decay}.
	}
	\label{fig_massspec}
\end{figure}

The mass distributions of the comoving number density of oscillons for $p=-0.5$ and $2$ at $\tilde \tau=81$ 
are shown in Fig.~\ref{fig_massspec} by the blue histograms.
Note that the vertical axis shows the number of oscillons in each bin normalized by $L^3$.
The orange and green regions show the oscillon distribution 
after the oscillon decay over $m\Delta t$
and will be discussed in Sec.~\ref{sec_timedep}.
Using these mass distributions, 
we calculate the comoving number density of oscillon $\tilde n_\tx{osc}$ 
and averaged mass $\braket{\tilde M_\tx{osc}} $ in Eq.~\eqref{eq_normalizedNum} for each $p$, which are
summarized in Table~\ref{table_results}.

\begingroup
\renewcommand{\arraystretch}{1.3}
\begin{table}[t]
	\caption{
	The results of simulations. 
	$\tilde \phi_{i,m}$ is the initial field value for efficient oscillon formation discussed in Sec.~\ref{sec_initialdependence}.
	$r_\tx{osc}$, $\tilde n_\tx{osc}$ and $\braket{\tilde M_\tx{osc}}$ are the oscillon energy ratio, number density, and averaged mass 
	when $\tilde \phi_{i} = \tilde \phi_{i,m}$.
	}
	\vspace{0.2cm}
	\centering
	\begin{tabular}{|c||c|c|c|c|} \hline
		$p$ & $\tilde \phi_{i,m}$ & $r_\tx{osc}$ &$	\tilde n_\tx{osc} $ & $\braket{\tilde M_\tx{osc}}$ \\ 
		\hline \hline
	    $-0.75$ & $12.0\pi$ & 0.73 & 0.98 & 2100 \\ \hline 
		$-0.5$  & $5.0\pi$  & 0.69 & 0.83 & 570 \\ \hline
		$0.5$   & $1.5\pi$  & 0.59 & 0.78 & 130 \\ \hline  
		$2$     & $0.7\pi$  & 0.60 & 0.79 & 47 \\ \hline
		$3$     & $0.55\pi$ & 0.62 & 0.85 & 32 \\ \hline
	\end{tabular}
	\label{table_results}
\end{table}
\endgroup

\subsection{Oscillon Power Spectrum}
\label{sec_powerspec}

To validate the analytical estimation of the power spectrum Eq.~\eqref{eq_oscillonPowerSpec},
we numerically calculate the power spectrum after oscillon formation.
In this subsection, 
we only show the results for $p=-0.5$ in the large box setup 
with $\tilde \phi_{i} = 5\pi$ in three dimensions 
and $\tilde \phi_{i} = 2.5\pi$ in two dimensions,
but similar result is obtained for $p=-0.75$ and $2$. 

\begin{figure}
	\centering
	\includegraphics[width=1.000\textwidth ]{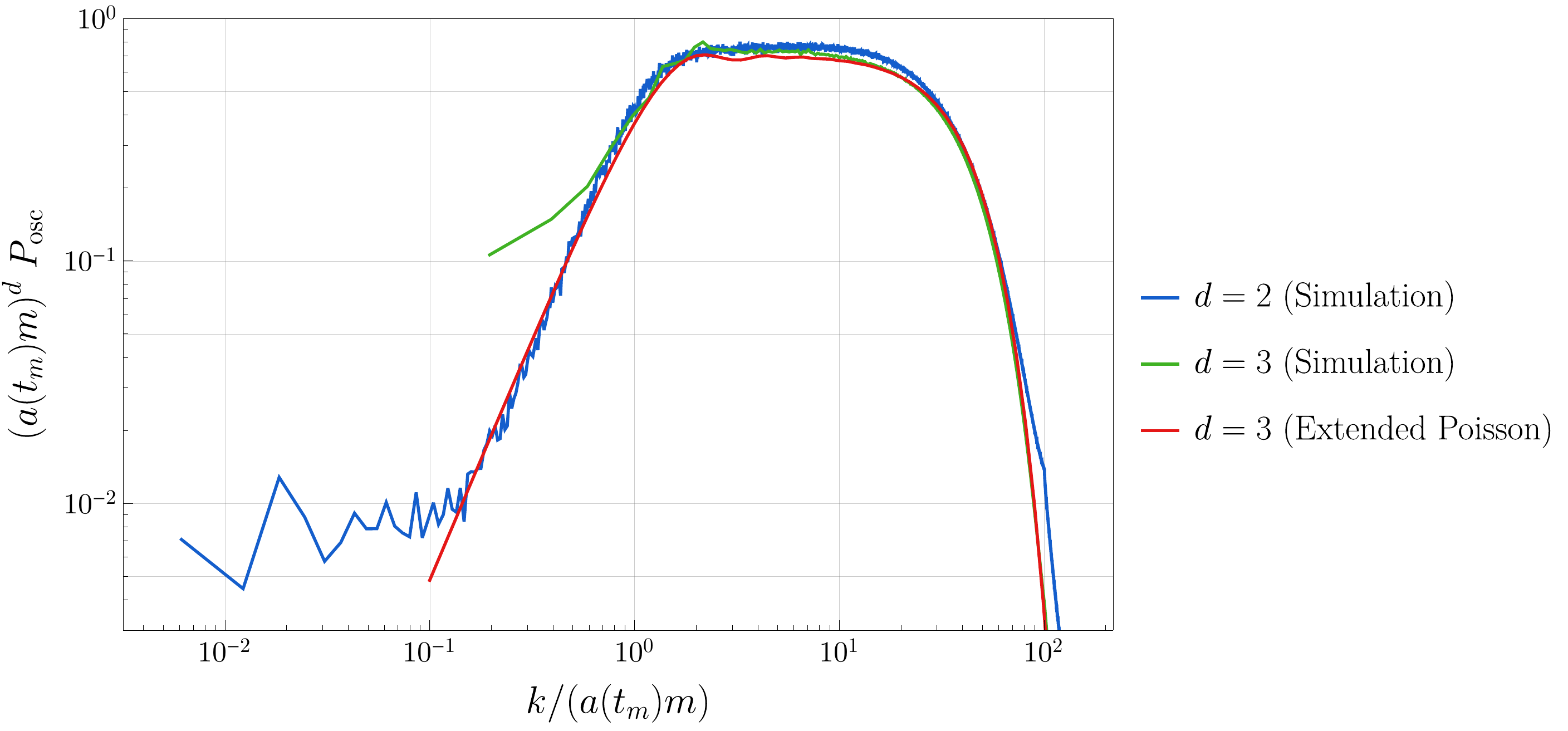}
	\caption{
		The normalized power spectrum with $p=-0.5$.
        The horizontal axis represents the normalized comoving wavenumber
        and the vertical axis represents the normalized power spectrum.
		We also plot the reproduced power spectrum based on the extended Poisson distribution discussed in Sec.~\ref{sec_poisson}.
	}
	\label{fig_PowerSpec}
\end{figure}

In Fig.~\ref{fig_PowerSpec},
the blue and green lines show the snapshot of the power spectrum 
at $\tilde \tau = 81.0$ when the oscillon formation has already finished.
The red line shows the analytical power spectrum from Eq.~\eqref{eq_oscillonPowerSpec} based on the oscillon distribution in Fig.~\ref{fig_massspec}.
When the cutoff scale $k_c$ in Eq.~\eqref{eq_oscillonPowerSpec} is taken as $k_c/(a(t_m) m) = 0.35$,
we find that the numerical power spectrum (green and blue lines) are well approximated by the analytical formula (red line).

The power spectrum consists of three parts.
The center plateau region corresponds to the Poisson distribution of the oscillons.
On the scale smaller than the oscillon size,
the power spectrum decreases following the oscillon profile given by Eq.~\eqref{eq_assumedprofile}.
On the scale larger than $k_c^{-1}$, 
the power spectrum decreases because of the energy conservation following Eq.~\eqref{eq_oscillonPowerSpecFactor}.
This suppression, however, does not hold for the lower end (see blue line in Fig.~\ref{fig_PowerSpec}).
One of the possibilities is that the oscillon energy ratio at each horizon itself fluctuates, 
which results in another source of Poisson noise with amplitude $\mathcal O(10^2)$ smaller than the plateau region.

\section{Time Evolution of Oscillons}
\label{sec_timedep}

As we introduced in Sec.~\ref{sec_oscillontheory}, 
the decay rate of the oscillons is too small to simulate numerically.
Therefore, in this section, 
we include the decay process of oscillons 
based on the analytic formula 
Eq.~\eqref{eq_oscillontimedepFormula}
to calculate the oscillon energy ratio $r_{\rm osc}$, 
the average oscillon energy $\langle M_{\rm osc} \rangle$, 
and the oscillon power spectrum.

First, 
we evolve the oscillon mass distribution derived in Sec.~\ref{sec_massdist}
by time $m\Delta t$
and the results of $p = -0.5$ and $2$ are shown in Fig.~\ref{fig_massspec}.
Since the lighter oscillons are more stable for $p=-0.5$, 
oscillons converge on the quasi-stable mass.
On the other hand, 
because for $p=2$ the oscillon decay rate has a pole at $M_\tx{osc}\sim 50$ (see Fig.~\ref{fig_decay})
where the decay rate is extremely small,
the oscillon distribution becomes spitted into two peaks
\footnote{
Note that the lifetime could be overestimated by poles for $p=0.5$, $2$ and $3$ as explained in Sec.~\ref{sec_oscillon_decay}.
}.

\begin{figure}
	\centering
	\begin{tabular}{c}
		
		\begin{minipage}{0.4\hsize}
			\begin{center}
					\includegraphics[width=1.\textwidth ]{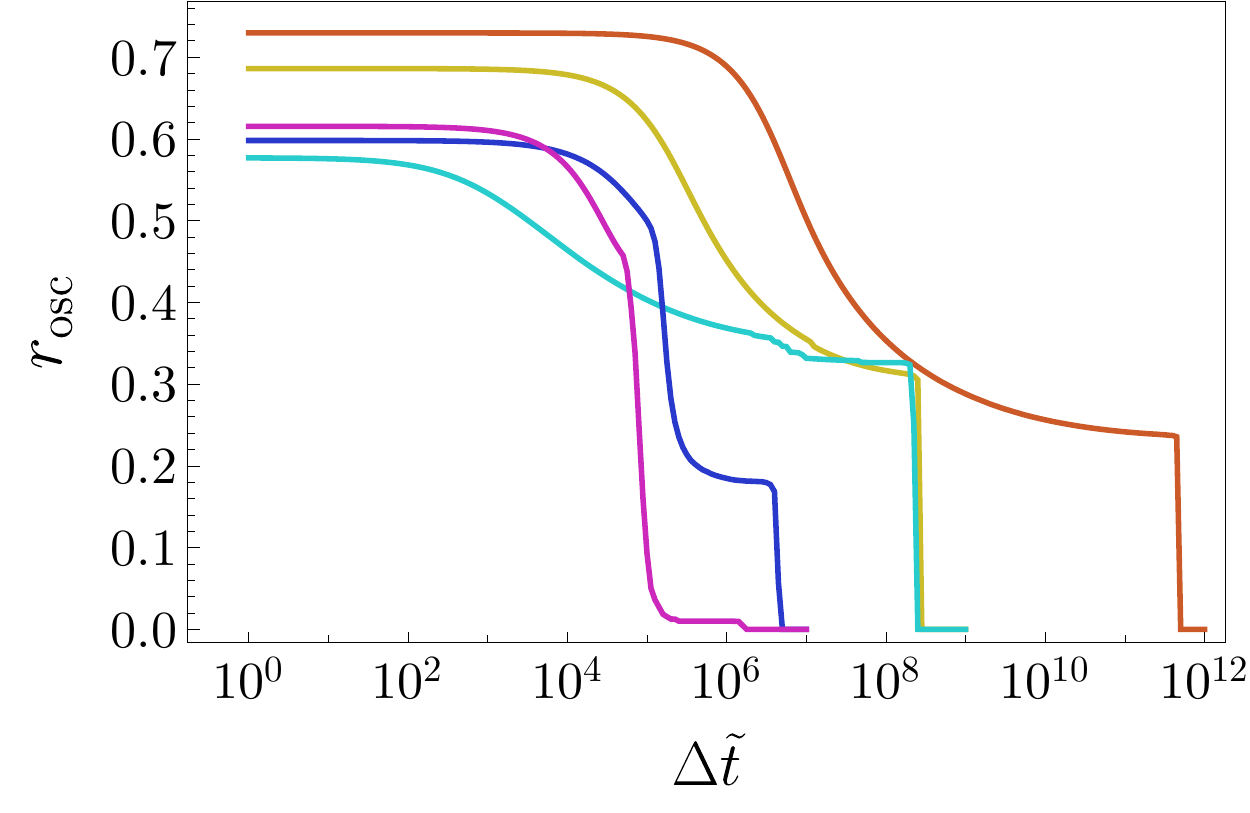}
			\end{center}
		\end{minipage}
		
		\begin{minipage}{0.51\hsize}
			\begin{center}
			 \includegraphics[width=1.\textwidth ]{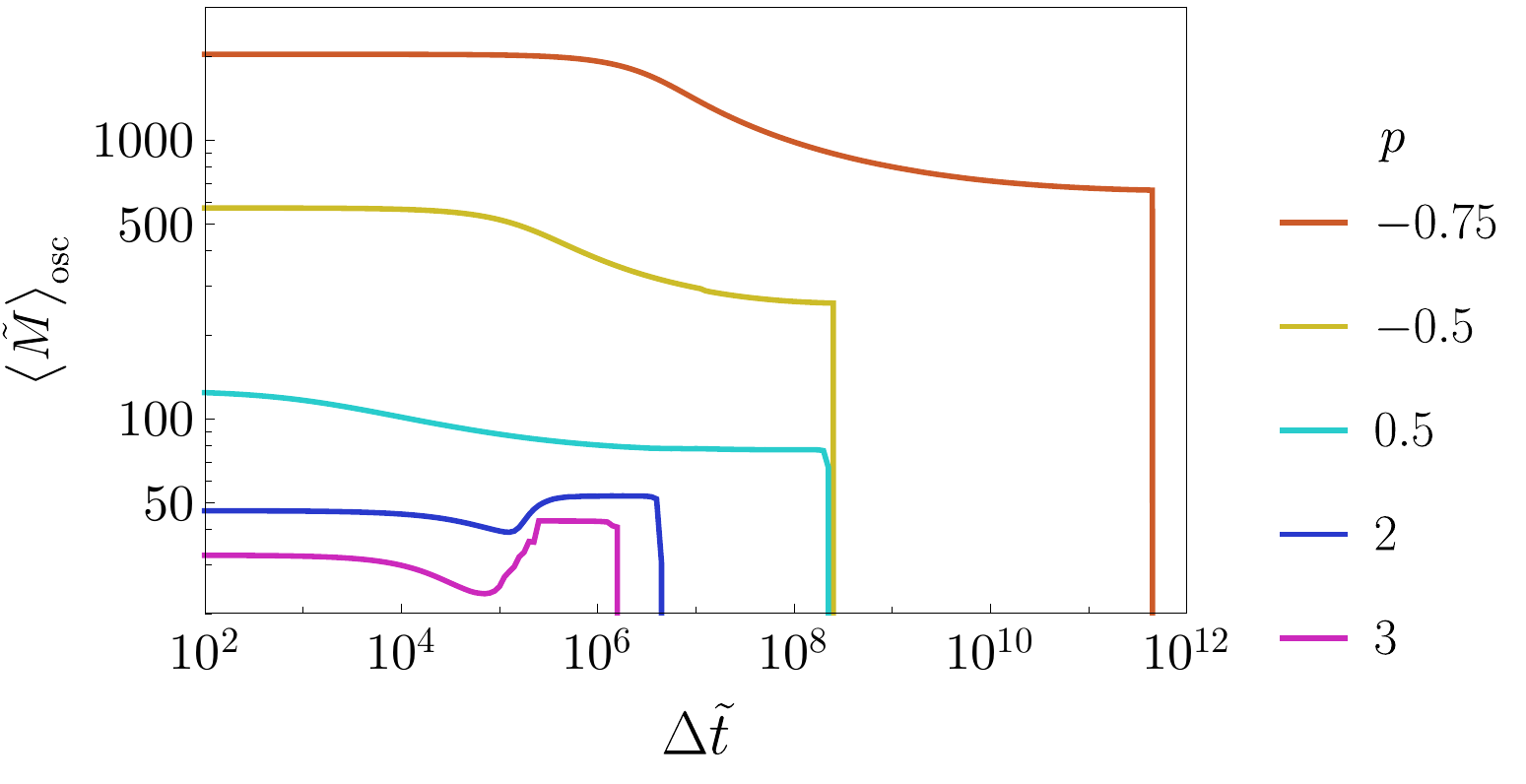}
			\end{center}
		\end{minipage}
		
	\end{tabular}
	\caption{
		The time dependence of the oscillon ratio and averaged mass when oscillons decay.
		We use the typical initial condition summarized in Table.~\ref{table_results}. 
		For $\Delta \tilde t \gtrsim 10^5$, the oscillons with $p=0.5,$ 2 and 3 are stabilized by the small decay rate at the pole, which has the uncertainty discussed in Sec.~\ref{sec_oscillon_decay}.
	}
	\label{fig_oscillondecay}
\end{figure}

Based on this mass distribution,
we calculate the time evolution of the oscillon ratio $r_{\rm osc}$ 
and the average oscillon mass $\braket{M_{\rm osc}}$ shown in Fig.~\ref{fig_oscillondecay}.
The time evolution of $\braket{M_{\rm osc}}$ is different for each $p$.
When $p=-0.75$ or $-0.5$, 
the average mass monotonically decreases 
because the lightest oscillon is the most stable 
as one can see from Fig.~\ref{fig_decay}.
On the other hand,
when $p= 0.5$, $2$ or $3$, 
the lifetime of oscillons at the poles of the decay rate is the longest.
Thus, the total number of oscillons $N_{\rm osc}$ decreases by the decay of smaller oscillons 
and it temporarily increases the average mass for $p=2$ and $3$ at $\Delta \tilde t\sim 10^5$ which corresponds to the lifetime of the smallest oscillon.

\begin{figure}[t]
	\centering
	\includegraphics[width=.75\textwidth ]{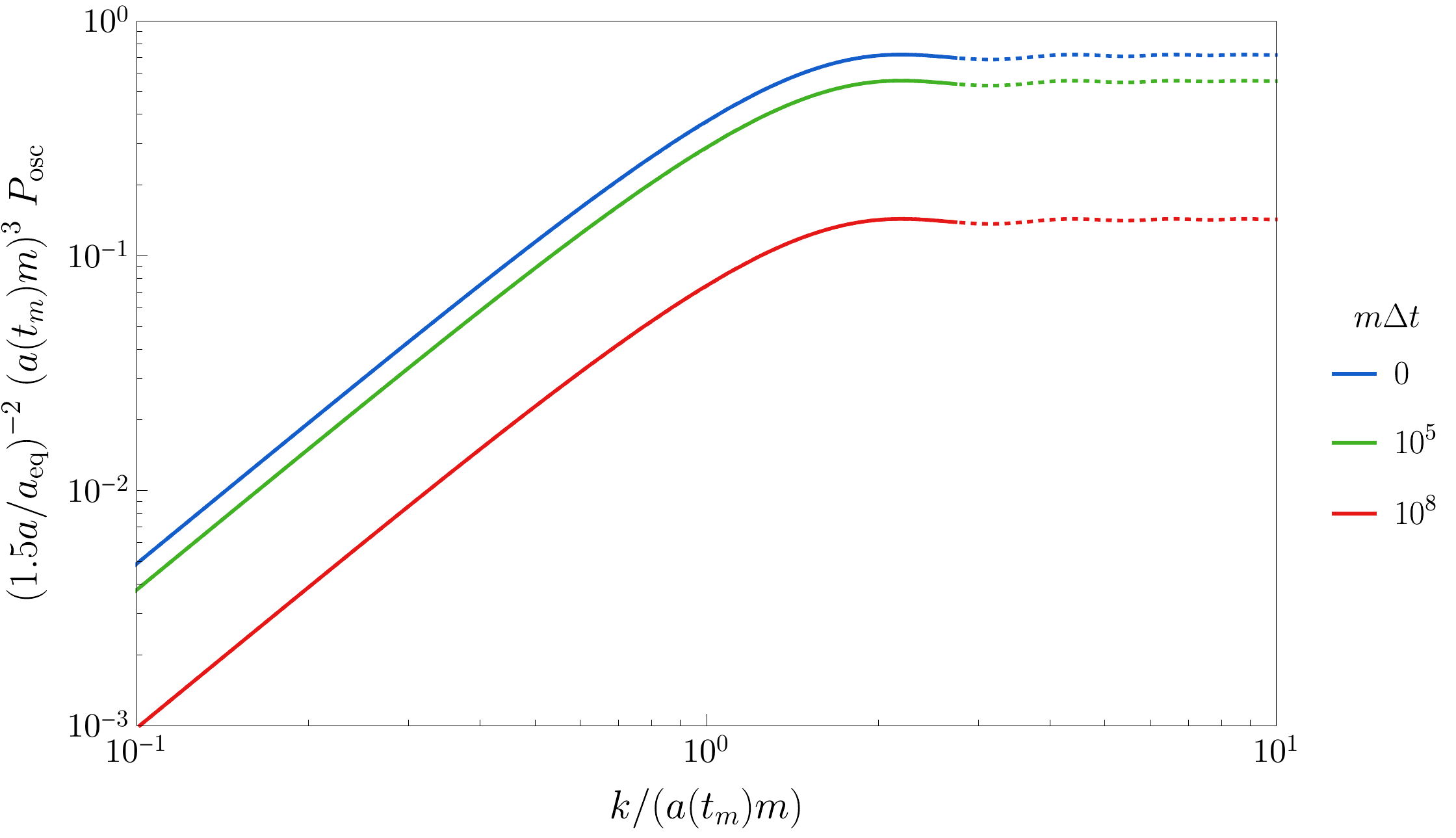}
	\caption{
		The power spectrum of oscillons for $p=-0.5$ based on Eq.~\eqref{eq_powerspec_real}.
		The different lines describe the power spectrum at different time including the effect of oscillon decay.
		We normalize the power spectrum so as to neglect the growth during matter dominated era.
		The dotted line shows the scales which includes less than 10 oscillons, $k > 2\pi ( n_\tx{osc}/10)^{1/3}$.
	}
	\label{fig_PowerSpec_timeevol}
\end{figure}

Based on these results, 
we consider the power spectrum after the partial decay of oscillons.
When the oscillon lifetime is long enough to survive until the matter-dominated era,
the power spectrum is gravitationally amplified.
Then, the realistic power spectrum during the matter-dominated era is represented by
\begin{align}
    P_{\tx{osc},k}(t) 
    &=
    \left( \frac{3}{2}\frac{a(t)}{a_\tx{eq}} \right)^2
	\frac{r_\tx{osc}^{2}}{n_{\rm osc} } 
	\frac{\Braket{M^2}}{\braket{M}^2}
	\left[
	1 - \left( \frac{2k_c}{k} \right)^2\sin^2\left( \frac{k}{2k_c} \right)
	\right]
	. 
	\label{eq_powerspec_real}
\end{align}
where $a_\tx{eq}$ is the scale factor at the matter radiation equality.
Note that the factor $\Braket{M^2}/\braket{M}^2$ converges on 1 
after the partial decay of oscillons because the mass distribution becomes monochromatic.
The result is plotted in Fig.~\ref{fig_PowerSpec_timeevol} for $p=-0.5$.
Because the gravitational growth only occurs on the scale 
which includes at least two oscillons,
we cut off the power spectrum on the scale
which includes 10 oscillons $n_{\rm osc} (2\pi/k)^3 > 10$
for instance, shown as the dotted lines.

Finally,
we comment on the power spectrum after oscillons have completely decayed.
Suppose that the oscillon with radius $R$ and energy $M$ breaks down 
radiating the ALPs with the kinetic energy $K \gtrsim m$
\footnote{
$K \gtrsim m$ is satisfied for the perturbative decay discussed in Sec.~\ref{sec_oscillon_decay}.
However, it is unclear whether this condition holds on the sudden decay at $dI/d\omega =0$, which is out of the scope of this paper.
},
the escape velocity from the oscillon surface 
$K_{\rm esc} \simeq G m M/R \sim m (F/M_{Pl})^2$
is smaller than $K$.
Thus, the decaying oscillon cannot gravitationally bind itself 
and the Poisson fluctuation would damp after the oscillon evaporation.

\section{Conclusion}
\label{sec_conclusion}

In this paper, 
we have performed a comprehensive analysis of oscillons
from its formation to the decay 
and have established the basics of oscillon phenomenology.

We firstly performed classical lattice simulation and calculated various quantities: 
the dependence of energy ratio of oscillons to the ALP field on the initial field value in Fig.~\ref{fig_roWithDiffPhi0},
the oscillon mass distribution in Fig.~\ref{fig_massspec},
and the power spectrum of oscillons in Fig.~\ref{fig_PowerSpec}.
We found that the oscillon formation becomes efficient
when the tachyonic instability which hinders the instability mode necessary for oscillon formation is weak.
Comparing the resultant power spectrum and the analytical formula derived in Sec.~\ref{sec_poisson},
we confirmed that the extended Poisson distribution is well approximation of the oscillon power spectrum.

We analyzed the time evolution of oscillons in Sec.~\ref{sec_timedep}
including the effect of their decay.
With the simulation results,
we calculated the current number density and typical mass of the oscillons.
Combining them with the oscillon decay and the gravitational growth,
we have estimated the power spectrum of oscillons 
after oscillons partially decay shown in Fig.~\ref{fig_PowerSpec}.

Finally, let us comment on the gravitational amplification of oscillon fluctuations.
Although we simply treated them as the matter fluctuations in Sec.~\ref{sec_timedep},
the evolution of the fluctuations produced by oscillons in the matter-dominated universe is still unclear.
Thus, 
given our result of the matter power spectrum as the initial condition,
we need a more precise study including the gravitational effect in future work.

\begin{acknowledgments}
We would like to thank T. Hiramatsu for useful comments on numerical calculations.
This work was supported by JSPS KAKENHI Grant Nos. 17H01131 (M.K.), 17K05434 (M.K.), 19H05810 (W.N.), JP19J21974 (H.N.), 19J12936 (E.S.), Advanced Leading Graduate Course for Photon Science (H.N.), and 
World Premier International Research Center
Initiative (WPI Initiative), MEXT, Japan (M.K.).
\end{acknowledgments}

\small
\bibliographystyle{JHEP}
\bibliography{Ref_ObjCount}

\appendix
\section{Suppression of Power Spectrum by Energy Conservation}
\label{app_energyConserve}

We estimate the effect of energy conservation, which suppresses the power spectrum on the large scale as Eq.~\eqref{eq_oscillonPowerSpecFactor}.
Here, we simply assume that all the energy of ALP is confined inside oscillons and oscillons have the same energy $M_o$.
In this case, Eq.\eqref{eq_oscillonmodes} contains the following sum of phases:
\begin{align}
    \psi =
    \sum_{i}^N\exp(-i\bs q_i\cdot\bs k ) ,
\end{align}
where index $i$ refers to $i$-th oscillon and $N$ is the total oscillon number.

At first, we revisit the calculation when the volume $V$ is smaller than the horizon size at oscillon formation, that is, we expect that the oscillons are randomly distributed.
Then, each phase factor $\bs x_i\cdot \bs k$ is a random variable, 
and the sum is estimated in the large number limit ($N\to \infty$) as
\begin{align}
	\left|
	\sum_{i}^N\exp(-i\bs x_i\cdot\bs k )
	\right|^2 
	= N.
\end{align}

Next, we include the effect of energy conservation.
Let us take $R$ as the scale for energy conservation where the energy transfer occurs only for scales smaller that $R$ 
and oscillon distribution is almost similar on the scales larger than $R$.
The oscillons are randomly distributed inside boxes with the volume $R^3$, 
that is, the position of a oscillon $\bs q =(q_x,q_y,q_z)$ has the uniform distribution $q_w \in [j_wR,(j_w+1)R]$ with $w=x$, $y$ and $z$. 
The index $\bs j = (j_x,j_y,j_z)$ refers to the different box in which the energy of oscillons is conserved independently.
The energy conservation requires that the number of oscillons in each box is universal constant $N_o$.
We take the total space with volume  $V=L^3$ and  calculate the power spectrum $\mathcal P_k$ with $k= 2\pi /L$, where $L$ is larger than the single box size, $J\equiv L/R\gg 1$.
Note that the total number of oscillons is now $N= N_o J^3$.

The average over the position of oscillons is given by 
\begin{align}
\braket{\cdots}_o
=\prod_{\bs j,n_j}
\left(
\int_{j_xR}^{(j_x+1)R} \frac{\df q_{x,n_j}}{R} 
\int_{j_yR}^{(j_y+1)R} \frac{\df q_{y,n_j}}{R} 
\int_{j_zR}^{(j_z+1)R} \frac{\df q_{z,n_j}}{R} 
\right) (\cdots),
\end{align}
where index $n_j$ refers to oscillons inside the $\bm j$-th box and $n_j$ runs from $1$ to $N_o$.
We take $\bm k = (2\pi/L,0,0)$ without loss of generality assuming the isotropic oscillon distribution.
The average of $\psi$ is calculated as 
\begin{align}
\braket{\psi}_o
&=\prod_{j_x' ,n_j'}
\left(
\int_{j_x'R}^{(j_x'+1)R} \frac{\df q_{n_j'}}{R} 
\right)
\sum_{j_x,n_j}^{J,N_o}\exp(-2\pi i q_{n_j}/L) ,
\\	&=
\sum_{j_x,n_j}^{J,N_o}
\frac{1}{-ikR}\left[
\exp(-2\pi i (j_x+1)/J) 
-\exp(-2\pi i j_x /J) 
\right] ,
\\	&=
N_o
\frac{\exp(-2\pi i/J)-1  }{-ikR}
\sum_{j_x}^{J}
\exp(-2\pi i j_x /J) 
=0,
\end{align}
where we use the orthogonal condition $	\sum_{j=1}^{J}\exp(-i2\pi j /J) =0$. 
The variance is calculated as
\begin{align}
\braket{|\psi|^2}_o
&=\prod_{ j_x'' ,n_j''}
\left(
\int_{j_x''R}^{(j_x''+1)R} \frac{\df q_{n_j''}}{R} 
\right)
\left|
\sum_{j_x,n_j}^{J ,N_o}\exp(-2\pi iq_{n_j}/L) 
\right|^2 ,
\\		&=\prod_{j'' ,n_j''}
\left(
\int_{j''R}^{(j''+1)R} \frac{\df q_{n_j''}}{R} 
\right)
\left[
\sum_{j, j'=1}^{J}
\sum_{n_j, n_{j'}'}^{N_o}e^{-2\pi i(q_{n_j} -q_{n_{j'}'}  )/L} 
\right],
\end{align}
where we omit the subscript "$x$" of $j$.
The average over $\df q_{n_j}$ is 1 for the terms with $j=j'$ and $n_j=n_{j'}'$ while the other combination can be integrated separately for $q_{n_j}$ and $q_{n'_{j'}}$ as
\begin{align}
\int_{j_xR}^{(j_x+1)R} \frac{\df q_{n_j}}{R}  
e^{-2\pi i q_{n_j} /L} 
=
\frac{	e^{-2\pi i/J}-1}{-ikR}
e^{-2\pi i j_x /J} 
=\epsilon e^{-2\pi i j_x /J} ,
\end{align} 
where we define the suppression factor:
\begin{align}
\epsilon = \frac{J}{-i2\pi }(e^{-2\pi i/J}-1).
\end{align}
The integration is performed as 
\begin{align}
\braket{|\psi|^2}_o
&=
N_o J 
+N_o(N_o-1)|\epsilon|^2J
+N_o^2|\epsilon|^2\sum_{j, j'(\neq j)}^{J}e^{-2\pi i (j-j' )/J} ,
\\&=
N_o J 
+N_o(N_o-1)|\epsilon|^2J
+N_o^2|\epsilon|^2
\left(
\left|\sum_{j=1}^{J}e^{-2\pi i j/J} \right|  ^2
-J
\right) ,
\\&=
N_o J (1-|\epsilon|^2)
=N\left[
1 - \left( \frac{2}{Rk} \right)^2\sin^2\left( \frac{Rk}{2} \right)
\right] .
\end{align}
For the modes smaller than the box size $Rk\gg1$, the variance coincides with the Poisson distribution, $\braket{|\psi|^2}_o\to N$.
On the other hand, for the modes larger than the box size $Rk\ll1$, the perturbation is suppressed by energy conservation.

\end{document}